# Resolving proximal nanometer objects below the diffraction limit with interferometric phase intensity nanoscopy

*Guangjie Cui[1#], Do Young Kim [1#], Di Zu[1], Guanbo Chai[1], Somin Eunice Lee[1]\**

[1]Department of Electrical & Computer Engineering, Biomedical Engineering, Applied Physics, Biointerfaces Institute, Macromolecular Science & Engineering, University of Michigan, Ann Arbor, Michigan USA
[#] Authors equally contributed.
\* To whom correspondence should be addressed. sleee@umich.edu

## Abstract

The ability to spatially and temporally map nanoscale environments in situ over extended timescales would be transformative for biology, biomedicine, and bioengineering. All nanometer objects, from nanoparticles down to single proteins, scatter light. Interferometric scattering stands as a powerful tool, offering ultrasensitivity and resolution vital for visualizing nanoscale entities. Interferometric scattering from an individual nanoparticle down to an individual protein has been detected; however, resolving adjacent nanometer objects with interferometric scattering has not yet been demonstrated. In this work, we present interferometric phase intensity nanoscopy to resolve adjacent nanometer objects with interferometric scattering. We demonstrate that multiphase and sensitivity of interferometric phase intensity nanoscopy reveals ellipse Airy patterns correlated with nanostructural features. We show that eliminating background fluctuation by employing circular polarized illumination in interferometric phase intensity nanoscopy is essential to separate proximal nanometer objects below the diffraction limit. We envision interferometric phase intensity nanoscopy for resolving a variety of adjacent nanometer objects from nanoparticles down to proximal proteins in situ over extended time periods for wide range of applications in biology, biomedicine and bioengineering.

**Introduction**

Mapping temporal (dynamic) heterogeneity of nanoenvironments (Figure 1a) is critical to design new smart materials and precision therapeutics, but broad timescales constrain time-resolved efforts at the expense of spatial resolution[1–4]. All nanometer objects, from nanoparticles down to single proteins, scatter light. Recently interferometric scattering has become a tremendously powerful optical imaging method with sensitivities capable of observing scattering from individual nanometer objects[5–17]. Due to the principle of interference, where the superposition of light waves can amplify or diminish their resultant intensity, interferometric scattering microscopy (iSCAT) detects the first order interference term rather than the quadratic scattering component, enabling detection of an individual nanoparticle and individual virus, and even an individual lipid and individual protein. As iSCAT utilizes scattering from the target, iSCAT is able to bypass photobleaching and phototoxicity problems occurring over extended time periods currently hampering traditional fluorescence microscopy.

To date, interferometric scattering from an individual nanoparticle down to an individual protein has been detected[5,6]; however, resolving adjacent nanometer objects in a diffraction limited region with interferometric scattering has not yet been demonstrated. While localization of an individual nanometer object using interferometric scattering is well established, resolving adjacent targets with interferometric scattering remains challenging. This is due in part because in coherent imaging, scattering light from adjacent targets no longer comes from the addition of intensity but the superposition of the amplitude and phase of each electro-magnetic field.

Here, we overcome these challenges by interferometric phase intensity nanoscopy. By integrating a phase-intensity multilayer thin film[18] with interferometric scattering, nanometer objects below the diffraction limit can be resolved. Scattering light from adjacent targets in coherent imaging is the superposition of the amplitude and phase of each electromagnetic field, but phase-intensity modulation in interferometric phase intensity nanoscopy enables to separate the intensity and phase such that nanometer objects are distinguished at different phases and adjacent targets can be then resolved. We have established that multiphase analysis and sensitivity of interferometric phase intensity nanoscopy bring to light ellipse Airy patterns that correlate with nanostructural characteristics. Our results also indicate that the use of circular polarized illumination in interferometric phase intensity nanoscopy is essential to break the diffraction threshold in differentiating between proximal nanometer objects. We envision interferometric phase intensity nanoscopy can resolve a variety of nanometer objects, from nanoparticles down to single proteins, in the future in order to map spatial and temporal (dynamic) heterogeneity of nanoenvironments in situ over extended timescales in biology, biomedicine and bioengineering.

**Results**

Interferometric scattering from an individual nanoparticle down to an individual protein has been detected; however, resolving adjacent nanometer objects with interferometric scattering has not yet been demonstrated. We resolve adjacent nanometer objects with interferometric scattering by interferometric phase intensity nanoscopy. Interferometric phase intensity nanoscopy provides an outstanding sensitivity beyond traditional interferometric scattering microscopy by remapping target's phase into intensity to reveal nanostructural details and resolve adjacent nanometer objects. Interferometric phase intensity nanoscopy detects the difference between a reference field and modulated sample scattered field.

$$I = |E_r + \mathcal{P}(E_S,\gamma)|^2 \approx |E_r|^2 - 2|E_r||\mathcal{P}(E_S,\gamma)|cos\phi \quad (1)$$

where $I$ is the intensity, $E_r$ is the reference field, $\mathcal{P}(E_S,\gamma)$ represents the modulated scattered field by PI, and $\phi$ denotes the relative phase between nanoprobes and reference plane irrelevant from modulation. The PI device in interferometric phase intensity nanoscopy plays an important role in modulate the intensity of scattered field:

$$|\mathcal{P}(E_S,\gamma)| = \left| \begin{bmatrix} 1 & 0 \\ 0 & 0 \end{bmatrix} \frac{1}{2} \begin{bmatrix} e^{i\frac{\gamma}{2}}+e^{-i\frac{\gamma}{2}} & e^{i\frac{\gamma}{2}}-e^{-i\frac{\gamma}{2}} \\ e^{i\frac{\gamma}{2}}-e^{-i\frac{\gamma}{2}} & e^{i\frac{\gamma}{2}}+e^{-i\frac{\gamma}{2}} \end{bmatrix} e^{i\frac{\pi}{4}} \begin{bmatrix} 1 & 0 \\ 0 & -i \end{bmatrix} \begin{bmatrix} E_S^x \\ E_S^y \end{bmatrix} \right| = \left| cos\frac{\gamma}{2}E_S^x + sin\frac{\gamma}{2}E_S^y \right| \quad (2)$$

where the terms of the Jones matrices represent a fixed quarter wave retarding element, variable full wave retarding element, and a fixed linear polarizing element of PI[18]. When the scattered light propagates through PI with phase retardation $\gamma$, the modulation of the contrast is characterized as:

$$C_{iPINE} = \frac{4|\mathcal{P}(E_S)|}{|E_r|} \cos\phi \qquad (3)$$

To implement interferometric phase intensity nanoscopy, we integrated PI into the detection path of an interferometric scattering microscope to achieve phase modulation (Figure 1b). Illumination laser (710 nm) was followed by a spatial filter and beam expansion optics to improve beam quality. In this work, we report eliminating background fluctuation by employing circular polarized illumination in interferometric phase intensity nanoscopy is necessary to separate adjacent nanometer objects below the diffraction limit. To convert illumination source from linear to circular, a quarter waveplate was inserted in the illumination path in order to eliminate signal to noise fluctuation and detector saturation during phase modulation. The major noise then is the shot noise which decreases with the increasing number of photons $N$ collected by the detector during the exposure time. Thus, the reference light amplitude remains constant to guarantee a high signal to noise as well as avoid signal to noise fluctuation and detector saturation.

$$\frac{d|\mathcal{P}(E_r,\gamma)|}{d\gamma} = \frac{\frac{1}{2}d|E_r|}{d\gamma} = 0 \qquad (4)$$

After reflecting on a beam splitter, the beam is focused on the back focal plane of the objective lens, creating a wide-field planar illumination on the sample. Both scattered light and reference beam are collected by the high numerical aperture oil immersed objective lens and propagated through PI, converting phase to intensity modulation. This phase to intensity modulation distinguishes interferometric phase intensity nanoscopy from traditional iSCAT to resolve adjacent nanoprobes below diffraction limit. As an example, four adjacent nanometer objects were indistinguishable in the diffraction limited interferometric scattering image (Figure 1c). Unlike iSCAT which only considers the intensity, interferometric phase intensity nanoscopy offers multiphase and sensitivity to reveal anisotropic Airy patterns correlated to nanostructural features. By employing modulation of different phases, interferometric phase intensity nanoscopy resolves adjacent nanometer objects below the diffraction limit (Figure 1c).

Notably, we observed elliptical Airy patterns by interferometric phase intensity nanoscopy which were not previously reported in traditional iSCAT[15,19]. We hypothesized that elliptical Airy patterns originate from the anisotropic nanostructural features of the target. The size of the elliptical Airy pattern would be limited by the diffraction limit, however, we anticipate the ellipticity and the intensity of the elliptical Airy pattern would vary due to the dependence between the scattered field amplitude and the scattering cross-section of the target. As a model candidate of an anisotropic [20–52] nanostructured target to validate the fundamental reason that we can see the ellipse Airy pattern, we simulated the scattered electric field pattern of gold nanorods (Figure 2a). Here, we acquired the nanostructural information from the ellipse Airy pattern by phase-intensity and phase-ellipticity. The phase-intensity reveals the eclipsed information (Figure 2b) from the background noise. Traditional iSCAT relies on differential images with and without the target of interest in order to reduce the background noise. However, interferometric phase intensity nanoscopy can detect the target which was initially covered by the background noise owing to the phase-intensity modulation. Figure 2b shows the detection of the target hidden by background noise. The target cannot be detected at $\gamma$ denoted by the purple shading because the contrast of target (gold nanorod; gold line) overlaps with the background noise (black line). On the other hand, at $\gamma$ denoted by the green shading, the contrast of target (gold nanorod, gold line) is significantly higher than the background (black line). Owing to the dependence between the scattered electric field and the scattering cross-section, the target's detectable phase range also varies. We varied the target size and observed larger targets show a higher contrast, allowing for a larger range of phase distribution and more ellipticity variation. (Figure 2c) Furthermore, not only is more data be collected, but interferometric phase intensity nanoscopy also acquires more nanostructural information than traditional iSCAT[15]. The simulation of the phase-intensity with the phase-ellipticity demonstrates the trend to the elliptical Airy pattern of each specific phase by PI. The scattered electric field vector finally makes some rotation and the reduction of the amplitude, which leads to ellipticity variation with the rotation of gold nanorods.

$$I_{det} = \cos^2\frac{\gamma}{2}|E_{r_x}|^2 + \cos^2\left(\frac{\gamma}{2} - \theta_p\right)|E_{s_x}|^2 + 2\cos\left(\frac{\gamma}{2} - \theta_p\right)\cos\frac{\gamma}{2}|E_{r_x}||E_{s_x}|\cos\Delta\varphi_x +$$
$$\sin^2\frac{\gamma}{2}|E_{r_y}|^2 + \sin^2\left(\frac{\gamma}{2} + \theta_p\right)|E_{s_y}|^2 + 2\sin\left(\frac{\gamma}{2} + \theta_p\right)\sin\frac{\gamma}{2}|E_{r_y}||E_{s_y}|\cos\Delta\varphi_y \quad (5)$$

where $I_{det}$ is the intensity of detector plane, $\frac{\gamma}{2}$ is the phase retardation through PI, $\theta_p$ is the major axis of gold nanorods from the x-axis, $E_{r_x}$, $E_{r_y}$ are the x- and y- component of reflected electric field from the glass slide, $E_{s_x}$, $E_{s_y}$ are the x- and y- component of the scattered electric field from the gold nanorods, and $\Delta\varphi_x$ $\Delta\varphi_y$ are the phase different between scattered E-field and reflected E-field for each axis. The detector plane intensity is governed by the derived equation, and we can get the exact rotation and variation of ellipticity by theoretical simulation in agreement with experimental data (Figure 2d). We found the same patterns with diffraction limited size with varying intensity from different size and phase in both theoretical and experimental results. We also detected the more dramatic changes of ellipticity with the larger size of particles with similar trend through the phase change. Thus, we can extract not only the phase information of nanometer objects but also employ modulation of ellipticity with different phases which can be applied to the interferometric phase intensity nanoscopy for resolving adjacent nano-objects below the diffraction limit.

To resolve adjacent nanometer objects below the diffraction limit, we found that eliminating background fluctuation by employing circular polarized illumination in interferometric phase intensity nanoscopy was necessary to promote resolution. Circular polarized illumination, as a key part in interferometric phase intensity nanoscopy, provides constant-intensity reference field and enables efficient background elimination for high contrast (Eq. 3). To this end, we compared illumination sources: circularly polarized illumination as compared to linearly polarized illumination in interferometric phase intensity nanoscopy. In Figure 3a, circularly polarized illumination provides an invariable background input which is correlated with nanostructural features (dipole moment), generating stable, in-phase contrast output over the phase modulation range. On the contrary, linearly polarized illumination caused the background to fluctuate over the phase modulation range as shown in the contrast images shown in Figure 3a (ii), leading to low contrast in dark frames. In addition, the misalignment between the dipole moment with respect to linear polarization main axis distorts the phase of contrast output during modulation, lowering overall contrast and making subsequent loss of the ability to resolve adjacent nanometer objects (Figure 3b). In experiment, the stability of the background intensity as indicated in Figure 3c (i) endorses the possibility to compensate for the intensity loss during modulation progress by increasing the overall illumination power or exposure time to catch up the upper-bound photon number limit which is usually determined by detector's full well capacity or samples photo-damage threshold. On the other hand as shown in Figure 3c (ii), nanoprobes under linear polarization illumination are feebly limited in low SNR dilemma: the dark retardation region ($\gamma \in [0.5\pi, 1.0\pi]$) with low SNR has redundancy in accommodating photons to reduce shot noise while higher photon flux brings risk of vastly saturation in the bright retardation region ($\gamma \in [1.5\pi, 2.0\pi]$). Here, we demonstrate that experimentally we can resolve adjacent nanometer objects. Figure 3d demonstrates the experiment of interferometric phase intensity nanoscopy in resolving adjacent nanoprobes with the approximate distance of 190 nm (diffraction limit: 380 nm for coherent imaging). Images were captured over the phase modulation range, averaging images at each $\gamma$ to repress shot noise. To remove the unwanted speckles from the distortion of beam wavefront, we modulated the stage with a frequency of 17 Hz at 2 cm/s. Median background image was calculated from a bundle of background images under each $\gamma$. Adjacent nanometer objects were indistinguishable in the diffraction limited image whereas adjacent nanometer objects were resolved in the interferometric phase intensity nanoscopy image (Figure 3d). Furthermore, interferometric phase intensity nanoscopy is not limited to the simple case of two nanoprobes. In Figure 3e, we extended the application of interferometric phase intensity nanoscopy to multiprobe conditions. Results show interferometric phase intensity nanoscopy can resolve multiple nanometer objects regardless of the more complicated interference introduced by the overlapping Airy patterns of adjacent nanoprobes (Figure 3e).

**Discussion**

In this work, we demonstrate the breakthrough of diffraction limit with interferometric phase intensity nanoscopy enabled by phase intensity modulation with circular polarized illumination. Some limitations in our current setup are nanoprobes have different responses to the phase intensity modulation of interferometric phase intensity nanoscopy. It is possible other nanometer objects could reject scattering circular polarized illumination light, making it difficult to detect and resolve them. As a solution, other modulation devices[44] could be inserted into the illumination optical path to replace the quarter waveplate enabling more complex modulation of the illumination source. Future work is underway to enable interferometric phase intensity nanoscopy to detect and resolve a wide variety of nanometer objects from nanoparticles down to single molecules.

**Conclusion**

In this article, we report interferometric phase intensity nanoscopy, which allows us to acquire nanostructural information from the ellipse Airy patterns and the varying phase intensity. Additionally, interferometric phase intensity nanoscopy enables to break the diffraction limit by mitigating the background fluctuation by PI, enabling sufficiently high contrast to address the nanoscale objects in close proximity. We anticipate that interferometric phase intensity nanoscopy would play critical roles in long time super-resolution applications which can reveal in situ observations over extended time periods in living cells down to single proteins in biology, biomedicine, and bioengineering.

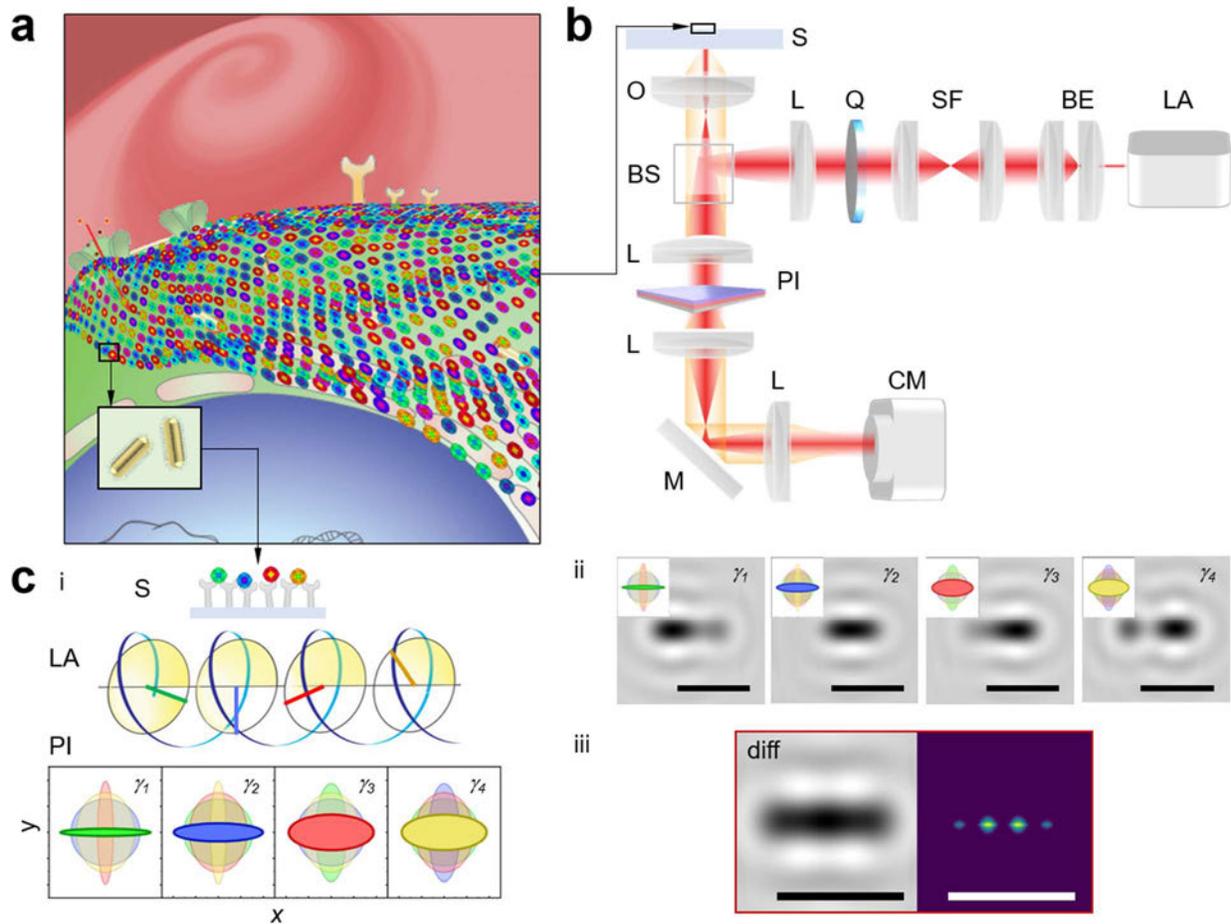

**Figure 1. Interferometric phase intensity nanoscopy - revealing nanostructural features - resolves adjacent nanometer objects below the diffraction limit. (a)** All nanometer objects, from nanoparticles down to single proteins, scatter light. In this work, adjacent nanometer objects, gold nanorods (Au), are resolved by interferometric phase intensity nanoscopy. We envision resolving adjacent proteins in living cells by interferometric phase intensity nanoscopy in the future. **(b) Experimental setup:** PI device, consisting of a fixed quarter wave retarding polymer film, a variable full wave retarding polymer film and a fixed linear polarizing polymer film, was integrated into the detection path of an interferometric scattering microscope to achieve phase modulation. In order to eliminate signal to noise fluctuation and detector saturation during phase modulation, quarter waveplate (Q) was inserted in the illumination path in order to convert the illumination source (LA) from linear to circular. S: sample, O: objective, L: lens, SF: spatial filter, BE: beam

expansion lenses, M: mirror, CM: CMOS camera. **(c)** **Principle of interferometric phase intensity nanoscopy:** i) Interferometric phase intensity nanoscopy resolves adjacent nanometer objects (green, blue, red, orange) below the diffraction limit. Nanometer objects are illuminated by circularly polarized light (LA). After phase modulation of $\gamma$ (PI), nanometer objects, diverse in their phases, exhibit different scattering electric fields. ii) Representative interferometric images ($\gamma_1$ = 0.41π rad, $\gamma_2$ = 0.88π rad, $\gamma_3$ = 1.12π rad, $\gamma_4$ = 1.82π rad) over the phase modulation range. (iii) Diffraction-limited interferometric scattering image. Interferometric phase intensity nanoscopy -resolved image. Scale bar: 500 nm.

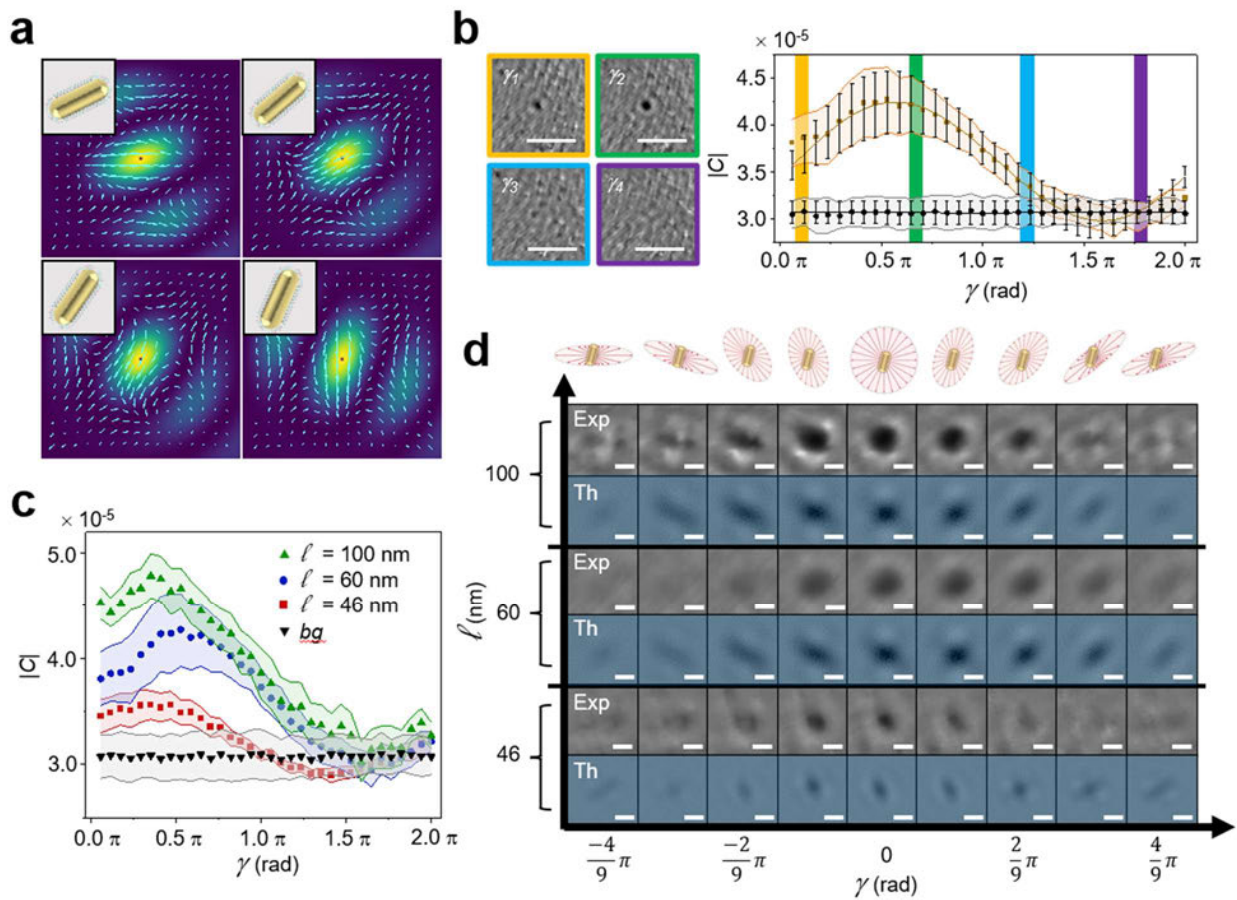

**Figure 2. Multiphase and sensitivity of interferometric phase intensity nanoscopy reveals ellipse Airy patterns correlated to nanostructural features. (a) Nanostructural feature (dipole moment)**: Far field radiation patterns, electric field vector plot (cyan arrows) overlaid with intensity pattern (color map) representing varying nanostructural feature (dipole moment) of model nanometer object (gold nanorod). **(b)** Owing to phase-intensity modulation, interferometric phase intensity nanoscopy distinguished the target which was hidden by the background noise. Representative interferometric images: $\gamma_1 = 0.1\pi$ rad (yellow shading), $\gamma_2 = 0.7\pi$ rad (green shading), $\gamma_3 = 1.2\pi$ rad (blue shading), $\gamma_4 = 1.8\pi$ rad (purple shading). Scale bar: $1\mu m$. Graph of absolute contrast |C| versus phase $\gamma$ for model nanometer object (gold nanorod), gold line and background noise, black line. **(c)** Graph of absolute contrast |C| versus phase $\gamma$ as a function of nanometer object size. **(d)** Interferometric phase intensity nanoscopy reveals ellipse Airy

**patterns:** Theoretical (Th) and experimental (Exp) interferometric scattering results over the phase modulation range $\gamma$ with respect to size. Scale bar: 200 nm.

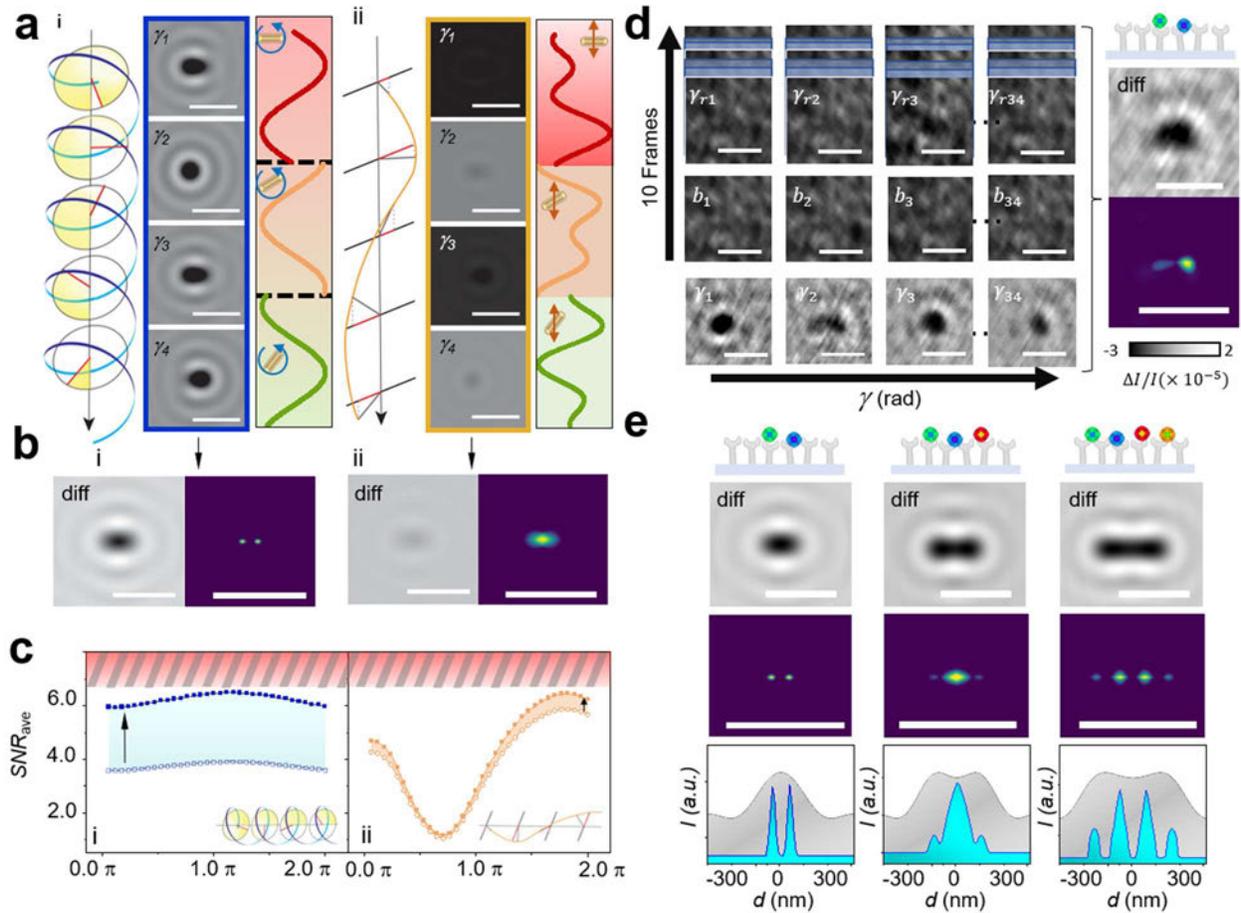

**Figure 3: Eliminating background fluctuation by employing circular polarized illumination in interferometric phase intensity nanoscopy is necessary to separate adjacent nanometer objects below the diffraction limit. (a)** i) Conceptual schematic of interferometric phase intensity nanoscopy$_c$: (left) input to PI, circularly polarized illumination, (middle) simulated interferometric images over the phase modulation range $\gamma$ from 0 to 2π, with interval of 0.4π, (right) output after PI where green, orange and red zones correspond to various dipole moments. ii) Conceptual schematic of interferometric phase intensity nanoscopy$_l$: (left) input to PI, linearly polarized illumination, (middle) simulated interferometric images over the phase modulation range $\gamma$ from 0 to 2π, with interval of 0.4π, (right) output after PI where green, orange and red zones correspond to various dipole moments. **(b) Interferometric phase intensity nanoscopy$_c$ is necessary to resolve proximal nanometer objects:** i) Interferometric phase intensity nanoscopy$_c$: (left)

diffraction-limited interferometric image, (right) interferometric phase intensity nanoscopy$_c$-resolved image. ii) interferometric phase intensity nanoscopy$_l$: (left) diffraction-limited interferometric image, (right) unresolved interferometric phase intensity nanoscopy$_l$ image. Scale bar: 500 nm. **(c)** Average signal to noise ratio graph of circular (left) and linear (right) polarized illumination over the phase modulation range. The shadowed zone on the top part of graph indicates the upper bound number of photons collected by detector. The circular polarized light (blue curve) holds a higher place in the SNR plot than linear polarized light (yellow curve) because more photons can be collected without saturation. Shot noise is considered here while the full well capacity of our sCMOS camera is 45000 $e^-$. **(d)** interferometric phase intensity nanoscopy$_c$ experimental procedure: Four out of thirty-four sets of images over the phase modulation range from 0 to 2π are shown. Target image is averaged from 10 frames under the same $\gamma$ to repress noise. While background image is generated by applying median filter to images captured when the stage is modulated. (top) Diffraction limited interferometric image. (bottom) interferometric phase intensity nanoscopy$_c$-resolved image. Scale bar: 500 nm. **(e) Interferometric phase intensity nanoscopy$_c$ resolves multiple proximal nanometer objects:** Simulation of two, three and four proximal nanometer objects. The distance between the adjacent nanoprobes is 150 nm while the diffraction limit is 300 nm in setup. Diffraction-limited interferometric image, interferometric phase intensity nanoscopy$_c$-resolved image. Scale bar = Scale bar: 500 nm. Intensity profile graphs plotted along line profiles in the corresponding diffraction limited interferometric image and interferometric phase intensity nanoscopy$_c$-resolved image. blue color: interferometric phase intensity nanoscopy$_c$-resolved. grey color: diffraction-limited.

# Acknowledgements

This work was supported by the Air Force Office of Scientific Research (AFOSR FA9550-19-1-0186, FA9550-22-1-0285, S.E.L.)